\documentclass[a4paper,twocolumn,11pt,accepted=2025-06-13]{quantumarticle}
\pdfoutput=1
\usepackage[numbers]{natbib}
\usepackage[utf8]{inputenc}
\usepackage[english]{babel}
\usepackage[T1]{fontenc}
\usepackage{amsmath}
\usepackage{hyperref}
\usepackage{amsfonts}

\usepackage{tikz}
\usepackage{lipsum}

\begin{document}

\title{Family of attainable geometric quantum speed limits}

\author{Zi-yi Mai}
\affiliation{School of Physics, Dalian University of Technology, Dalian
116024, P.R. China}
\orcid{0009-0005-7000-4247}
\author{Zheng Liu}
\affiliation{School of Physics, Dalian University of Technology, Dalian
116024, P.R. China}
\orcid{0009-0006-3322-1901}
\author{Chang-shui Yu}
\affiliation{School of Physics, Dalian University of Technology, Dalian
116024, P.R. China}
\orcid{0000-0002-8174-3775}
\email{ycs@dlut.edu.cn}
\maketitle

\begin{abstract}
  We propose a quantum state distance and develop a family of geometrical quantum speed limits (QSLs) for open and closed systems. The family of the QSL time includes an adjustable scalar function,  by which we derive three QSL times with the functions that are particularly chosen in this paper. Two of the derived QSL times are exactly the ones presented in Ref. \cite{Campaioli2019tightrobust} and \cite{mai2023tight}, respectively, and the third one can provide a unified QSL time for both open and closed systems. The three QSL times are attainable for any given initial state in the sense that dynamics exist, driving the initial state to evolve along the geodesic.  We numerically compare the tightness of the three QSL times, which typically promises a tighter QSL time if optimizing the function.
\end{abstract}

\section{Introduction}
Quantum speed limit (QSL) is a crucial feature of the quantum system; it presents a lower bound of required time for a dynamics process, which drives a given initial quantum state to the given target state. The QSL is an essential topic of interest in application scenarios, such as quantum optimal control \cite{Deffner_2017, PhysRevLett.103.240501, PhysRevA.92.062110, PhysRevLett.109.115703, PhysRevLett.111.260501, PhysRevA.82.022318, PhysRevLett.118.100601, PhysRevLett.118.100602}, quantum information processing \cite{PhysRevA.95.042314, PhysRevA.90.012303}, computational gates \cite{lloyd2000ultimate}, thermometry \cite{Campbell_2018}, quantum metrology \cite{giovannetti2011advances, PhysRevLett.96.010401, PhysRevLett.109.233601}, quantum thermodynamics \cite{PhysRevA.104.042202} and so on.

The most typical QSL is the  Mandelstam-Tamm (MT) bound \cite{Mandelstam1991} $\tau_{MT}^\bot=\pi/(2\Delta E)$  for a closed system with energy variance of $\Delta E$, which,  established by Mandelstam and Tamm, provides lower bound of the required time to drive a pure state to its orthogonal state by time-independent Hamiltonian.  Margolus and Levitin (ML) then give a different version of QSL $\tau_{ML}^\bot=\pi/(2E)$ \cite{MARGOLUS1998188} for unitary evolution between orthogonal states in terms of average energy $E$.  The combination of the above two bounds,  i.e., the MT-ML bound $\tau_{MT-ML}^\bot=\pi/(2\min\left\{E,\Delta E\right\})$,  is suggested for a tight bound \cite{Vittorio_Giovannetti_2004}.  It is shown that the MT-ML bound is attainable for the states superposed by two eigenstates of the Hamiltonian with the average energy equal to the energy variance  \cite{PhysRevLett.103.160502}.
Many studies have been carried out to generalize the concept of QSL bound to various cases, including the QSLs with incompletely distinguishable state pair \cite{K_Bhattacharyya_1983,fleming1973unitarity, PhysRevResearch.5.043234}, time-dependent Hamiltonian \cite{PhysRevLett.65.1697, PhysRevA.108.052421}, and tight bound \cite{PhysRevLett.129.140403, Campaioli2019tightrobust,PhysRevLett.120.060409,Hornedal_2022}.  Due to the interaction with environments,  QSL has also been generalized to open systems \cite{PhysRevX.12.011038, PhysRevA.93.052331, Deffner_2013,russell2014geometrical, PhysRevA.103.022210, PhysRevA.95.022115, MONDAL2016689,PhysRevA.95.052104, PhysRevA.98.042132, PhysRevLett.120.060409,zhang2014quantum, MONDAL20161395, Andersson_2014,PhysRevLett.110.050403,sun2015quantum, PhysRevLett.123.180403, PhysRevA.86.016101,PhysRevA.67.052109}. In addition,  the bound on QSL has demonstrated some physical meanings by linking to some physical quantities or resources such as entanglement \cite{PhysRevX.9.011034, PhysRevA.78.042305, Zander_2007}, coherence \cite{PhysRevA.93.052331, MONDAL2016689,Mohan_2022, PhysRevA.102.053716, PhysRevLett.118.140403, PhysRevLett.119.130401}, purity \cite{PhysRevResearch.2.023299} and entropy \cite{Campaioli_2022, Funo_2019}, etc.. Even though explicit physical meanings are expected for the QSLTs,  these QSLTs are often limited in quite particular systems or very loose, and most of their attainability in general cases hasn't been proved or at least is challenging to establish. How to present an attainable or tight QSLT for general systems remains a challenging question.

\begin{figure}[t]
  \centering
\includegraphics[width=0.48\textwidth]{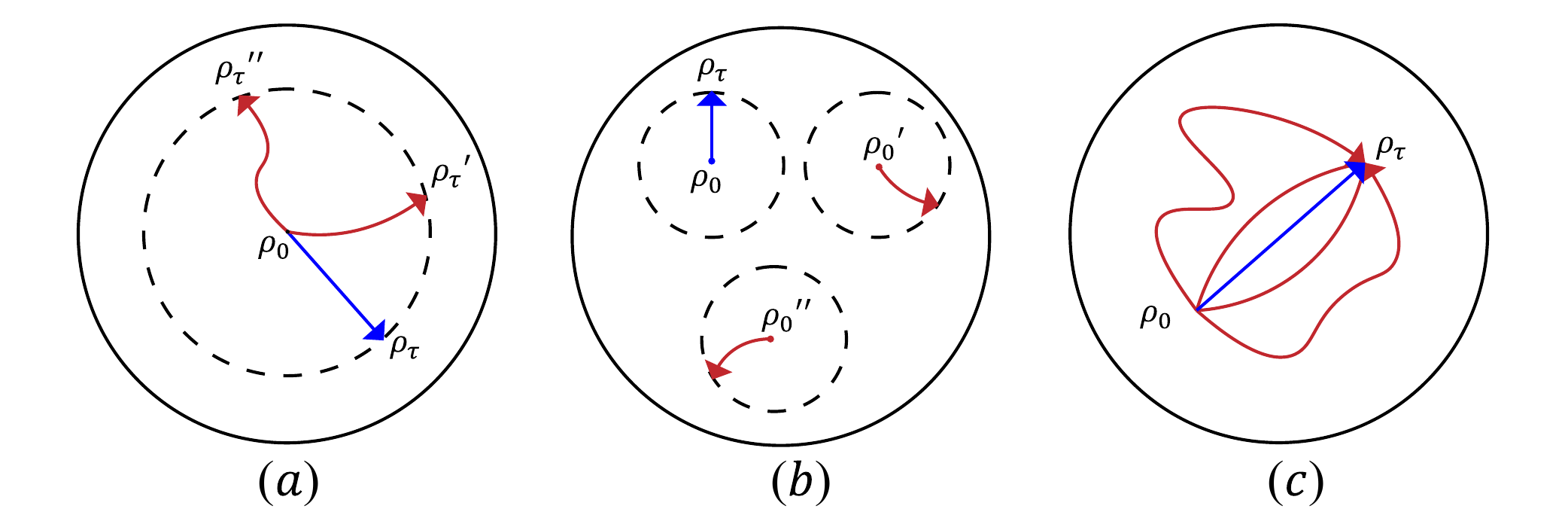}
\caption{Geometrical diagram of optimization problem of QSLs.  (a) Evolving a given distance from the given initial state $\rho_0$.  (b) Evolving a given distance within some constraints. (c)  Different dynamics connecting a given pair of $\rho_0$ and $\rho_\tau$.
}\label{qsl_fig}
\end{figure}

The essence of QSL is an optimization problem subject to some constraints, which has different understandings corresponding to various scenarios \cite{frey2016quantum}.   For example,   if a particular evolution trajectory is considered,  the QSL time (QSLT)  describes the deviation degree of evolution trajectory driven by the dynamics from the geodesic connecting its initial and final state (or the given initial state and final states if the initial state is fixed).  Additionally,  as is shown in Figure \ref{qsl_fig} (a),  if an evolution distance from a given initial state and some `energy scale' (usually the denominator of the  QSLT) are fixed, the QSLT characterizes the minimum time to travel this distance from the given initial state.  Based on Figure 1 (a), if the initial state is not emphasized,  the QSLT means the minimum time to travel the claimed distance from any initial state, which is sketched in Figure 1 (b).  If the initial and final states are fixed, the QSLT characterizes the minimal evolution time required for dynamics to connect the two states, as is shown in Figure 1 (c).    The optimization implies infinitely many dynamics need to be considered, so the evolution time is larger than the QSLT if the dynamics are not optimal.  In addition, infinitely many QSLTs can also be induced based on the different state distances.   From these perspectives, needless to cover the physical meaning, the tightness and the attainability of a QSL from a mathematical perspective are also challenging to deal with.  In this sense, it is naturally expected to establish a QSLT as tight as possible or even an attainable QSLT.

In this paper,  we establish a family of attainable geometrical QSLTs for open systems based on our developed distance measure of quantum states, which, as a vital ingredient of the QSLT,  is constructed by redefining the geometry of quantum states.  A  prominent feature of the quantum state geometry is the dependence on a function, which can induce a large family of QSLTs.   Maximizing the function $f$ typically gives a QSLT, which can reduce to the bound given in Ref.  \cite{PhysRevLett.120.060409} for unitary dynamics.  In addition, we mainly consider three particular choices of the function $f$ and obtain the corresponding QSLTs, which indicates that two QSLTs can reduce to the previous QSLTs and the third one provides a unified QSLT  of a two-level system for unitary and non-unitary evolution. Namely, it can be saturated by both the unitary and nonunitary dynamics. It reveals that the unitary and nonunitary dynamics can drive the two-level system to evolve along the geodesics. This can be seen as the optimal evolution trajectory connecting two points.  All three QSLTs are attainable for open systems because any initial state can evolve along the geodesic, as shown in Fig. \ref{qsl_fig} (a); in particular, the third QSLT is saturated by pure depolarizing dynamics.   This paper is organized as follows. We first present the state map, define the distance of density matrices,  and then establish a family of function-dependent QSLTs.  Then, we explicitly study three QSLTs by choosing a particular function and demonstrating attainability through specific dynamics.  Finally, we take numerical examples of dynamics to discuss the QSLT's tightness.

\section{The distance and the geometric QSL}
\begin{figure}[t]
  \centering
\includegraphics[width=0.48\textwidth]{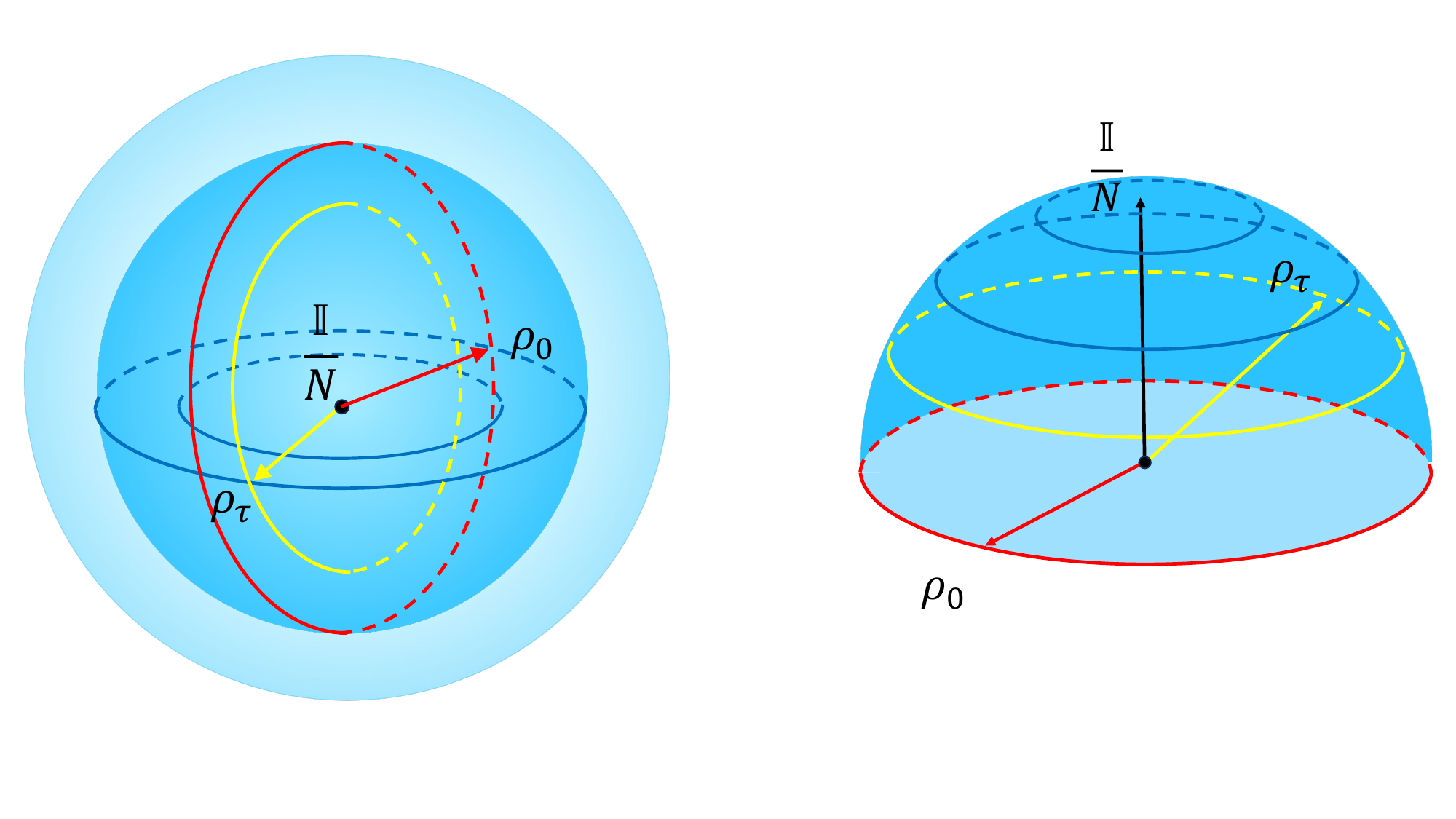}
\caption{The  Bloch sphere (left) and the (partly) metric space $F(\mathcal{D})$ (right) for a qubit state.   In the Bloch sphere, the red and yellow vectors indicate the initial and final states for a physical process, respectively, and the outermost blue outline indicates the outer surface of the Bloch sphere, i.e., the set of pure states. The metric space (right) is equipped with a distance function (\ref{D}). Here we take $f(x)=\mathrm{Tr}\rho_0^2$ for example.  In this case,  all the states with the purity $\mathrm{Tr}\rho_0^2$ are located at the equator.   In fact, the right spherical cap only corresponds to the states located in the cross-section determined by $\rho_0$ and $\frac{\mathbb{I}}{N}$ in the left Bloch ball.  The total state space of $F(\mathcal D)$ is the spherical caps collection.  Here we only take a qubit system as an example,  which can be generalized to high-dimensional system in principle, but cannot be given in such an intuitive way.
}\label{geometry_fig}
\end{figure}
\textit{The distance.-}To begin with, we first construct a $N$-dimensional metric space for the quantum state similar to the classical Bloch representation.
Let $f(x)$ be a function of scalar $x\in[\frac{1}{N},1]$ such that $f(x)\geq x$, $\frac{d}{dx}f(x)\leq 1$ and  $\sum_{i=1}^2\left\vert f(x_i)-x_i\right\vert\neq 0$ for $x_1\neq x_2$. For convenience, we denote the set consisting of all the functions $f$ satisfying the above constraints as $\mathcal F$. Let $\mathcal M$ denote the set of all the Hermitian matrices and $\mathcal D$ denote the set of all the positive density matrices.  Given a density matrix $\rho\in \mathcal D$,  define a map $F:\mathcal D\rightarrow \mathcal M$ as
\begin{equation}\label{f}
  F(\rho)=\begin{cases}
            \frac{\mathbb I}{\sqrt{N}}, & \mbox{if } \rho=\frac{\mathbb I}{N} \\
            \frac{\rho+(\sqrt{N}\sqrt{f(\mathrm{Tr}\rho^2)-\mathrm{Tr}\rho^2}-1)\mathbb{I}/N}{\sqrt{f(\mathrm{Tr}\rho^2)-1/N}}, & \mbox{otherwise},
          \end{cases}
\end{equation}
where $\mathbb{I}$ is the identity operator.
It is evident that for a particular function $f(x)\in\mathcal F$, $F(\rho)$ is uniquely determined for a given $\rho$. Verifying that $F(\rho)$ is continuous is not difficult.
Consider the  Hilbert-Schmidt inner product, i.e.,  $\langle A, B\rangle=\mathrm{Tr}A^\dagger B=\mathrm{Tr}AB$ for any pair of Hermitian matrices $A, B$.
One can find that  $-1\leq \langle F(\rho),F(\sigma)\rangle\leq 1$ for any pair of density matrices $\rho$ and $\sigma$ and   $\langle F(\rho),F(\sigma)\rangle= 1$ $\Leftrightarrow$ $\rho=\sigma$. Thus, we can define the distance of two density matrices as follows.

\textit{Definition 1.-} Given any two density matrices $\rho$ and $\sigma$,  their distance is defined as
\begin{equation}\label{D}
  D(\rho,\sigma)=\arccos \langle F(\rho),F(\sigma)\rangle
\end{equation}
with  $D(\rho,\sigma) =0$ iff $\rho=\sigma$.

The above map indicates that each density matrix is mapped to a unit sphere in the Hermite matrix space $\mathcal M$ equipped with a Hilbert Schmidt norm.  We have proved that this mapping $F$ is injective in Appendix B,  so $F$ essentially identifies each density operator using a unique point on the unit sphere. Then, the distance between the two density operators can be defined as the distance between the corresponding points on the sphere, which happens to be Eq. (\ref{D}). Therefore, Eq. (\ref{D}) is a well-defined distance metric. Figure. \ref{geometry_fig} can provide an intuitive understanding of the metric space. It can be seen that $\left\langle F\left(\mathbb I/N\right), F\left (\rho\right)\right\rangle=\sqrt{\frac{f(\mathrm{Tr}\rho^2)-\mathrm{Tr}\rho^2}{f(\mathrm{Tr}\rho^2)-\frac{1}{N}}}$, the intersection angle of $F(\mathbb I/ N)$ and $F(\rho)$ is entirely determined by the purity of $\rho$.  Considering the function $f(x)\in\mathcal F$, which leads to a fact that $\left\langle F\left(\mathbb I/N\right), F(\rho)\right\rangle$ is monotonically decreasing with respective to the purity of the density matrix due to $\frac{d}{dx}f(x)\leq 1$ and $f(x)\geq x$. It can be shown that
\begin{eqnarray}\label{deri_dis1}
  &\frac{d}{d\mathrm{Tr}\rho^2}\left\langle F\left(\frac{\mathbb I}{N}\right),F(\rho)\right\rangle=\left[\frac{d}{dx}\sqrt{\frac{f(x)-x}{f(x)-\frac{1}{N}}}\right]_{x=\mathrm{Tr}\rho^2}\\
  &=\left[\frac{\frac{d}{dx}f(x)(x-\frac{1}{N})-(f(x)-\frac{1}{N})}{2\sqrt{f(x)-x}\left(f(x)-\frac{1}{N}\right)^{\frac{3}{2}}}\right]_{x=\mathrm{Tr}\rho^2}\leq 0,
\end{eqnarray}
where the final inequality is obtained by using $\frac{d}{dx}f(x)\leq 1\leq (f(x)-1/N)/(x-1/N)$.
That is, the density matrix with less mixedness shows a larger intersection angle with the maximally mixed state. In this sense, the metric space $F(\mathcal D)$ is a unit spherical cap whose north pole is the maximally mixed state and whose latitudes consist of the density matrices with the same purity. It also indicates that $\left\langle F(\mathbb I/N), F(\rho)\right\rangle$ labels the latitude related to the purity $\mathrm{Tr}\rho^2$, and Eq. (\ref{deri_dis1}) indicates that density matrices with less purity occupy the latitude which is closer to the north pole, which is a natural result because one can decrease the purity of the system by mixing the maximally mixed state. The function $f\in\mathcal F$ adjusts the distribution of the latitudes with different purities. For an example of the unitary dynamics,  let $f(x)=\mathrm{Tr}\rho_0^2$ as a constant for a fixed $\rho_0$.  In Figure \ref{geometry_fig}, the latitude with the purity of $\mathrm{Tr}\rho_0^2$ coincides with the equator of the semi-sphere.

\textit{The QSLT.-} Now let's consider a density matrix $\rho_t$ at the moment $t$ evolves an infinitesimal duration $dt$, the final state can be given as $\sigma=\rho_t+d\rho_t$ with
$d\rho_t=dt\dot{\rho}_t$ determined by its dynamics. The distance between the initial state $\rho_t$ and the final state $\rho_{t+dt}$ can be written as $ds=D(\rho_t,\rho_t+d{\rho}_t)$, or equivalently,
\begin{widetext}
\begin{equation}
\cos D(\rho_t,\rho_t+d{\rho}_t)
=1-\frac{1}{2}ds^2
  =\frac{\mathrm{Tr}\rho_t(\rho_t+d\rho_t)-1/N+\sqrt{f_t-\mathrm{Tr}\rho_t^2}\sqrt{f_{t+dt}-\mathrm{Tr}\rho_{t+dt}^2}}{\sqrt{f_t-1/N}\sqrt{f_{t+dt}-1/N}},\label{s0}
\end{equation}
\end{widetext}
  where $f_t=f(\mathrm{Tr}\rho_t^2)$.
  One can always expand  the function $f_{t+dt}=f(\mathrm{Tr}(\rho_t+d\rho_t)^2)$ to the second order  as
\begin{equation}
\begin{split}
   f_{t+dt}=&f_t+2(f')_t\mathrm{Tr}\rho_td\rho_t+(f')_t\mathrm{Tr}(d\rho_t)^2\\
   &+{2(f'')_t(\mathrm{Tr}\rho_td\rho_t)^2},\label{c1}
\end{split}
\end{equation}
 where $(f')_t$ and  $(f'')_t$ denotes the first-order derivative  $\frac{df(x)}{dx}\Big\vert_{x=\mathrm{Tr}\rho_t^2}$  and the second-order derivative  $\frac{d^2f(x)}{dx^2}\Big\vert_{x=\mathrm{Tr}\rho_t^2}$, respectively.
Similarly, one can find that
\begin{equation}
\frac{1}{\sqrt{f_{t+dt}-1/N}}=\frac{1-\frac{\Delta}{2(f_t-1/N)}+\frac{3\Delta^2}{8(f_t-1/N)^2}}{\sqrt{f_t-1/N}}\\ \label{c2}
  \end{equation}
  and
  \begin{equation}
  \begin{split}
  &\sqrt{f_{t+dt}-\mathrm{Tr}(\rho_t+d\rho_t)^2}=\sqrt{f_t-\mathrm{Tr}\rho_t^2}\times\\
  &\left(1+\frac{\delta}{2(f_t-\mathrm{Tr}\rho_t^2)}-\frac{\delta^2}{8(f_t-\mathrm{Tr}\rho_t)^2}\right) ,\label{c3}
  \end{split}
  \end{equation}
  where
  \begin{widetext}
  \begin{equation}
  \Delta=2(f')_t\mathrm{Tr}\rho_td\rho_t+(f')_t\mathrm{Tr}(d\rho_t)^2
  +2(f'')_t(\mathrm{Tr}\rho_td\rho_t)^2
  \end{equation}
  \begin{equation}
  \delta=2\left[(f')_t-1\right]\mathrm{Tr}\rho_td\rho_t+\left[(f')_t-1\right]\mathrm{Tr}(d\rho_t)^2
  +2(f'')_t(\mathrm{Tr}\rho_td\rho_t)^2
  \end{equation}
\end{widetext}
  are both determined by Eq. (\ref{s0}), where Eq. (\ref{c3}) is obtained by using the Taylor expansion $\sqrt{X+dx}=\sqrt{X}\left(1+\frac{dx}{2X}-\frac{dx^2}{8X^2}\right)$.
Substituting Eqs. (\ref{c1},\ref{c2},\ref{c3}) into Eq. ({\ref{s0}}), one will immediately  obtain the metric as
\begin{widetext}
\begin{equation}\label{ds}
  ds=\sqrt{\frac{\mathrm{Tr}(d\rho_t^2)+\left(\frac{(f')_t\left(2-3(f')_t\right)}{f_t-\frac{1}{N}}+\frac{\left((f')_t-1\right)^2}{f_t-\mathrm{Tr}\rho_t^2}\right)(\mathrm{Tr}\rho_td\rho_t)^2}{f_t-\frac{1}{N}}}.
\end{equation}
\end{widetext}

Consider a dynamic evolution of a density matrix $\dot{\rho}_t$ from an initial state $\rho_0$ to the final state $\rho_\tau$, based on the triangle inequality of the distance, we have $D(\rho_0,\rho_\tau)\leq \int_0^\tau ds$.  Define the  average evolution speed  as $ \bar{v}=\frac{1}{\tau}\int_0^\tau {ds}$,  then  the QSLT can be given as
\begin{eqnarray}\label{QSL}
& \tau_{qsl}^f= \frac{D(\rho_0,\rho_\tau)}{\bar{v}}=\frac{D(\rho_0,\rho_\tau)}{\frac{1}{\tau}\int_0^\tau \frac{ds}{dt}dt},
\end{eqnarray}
where
\begin{equation}\label{dsdt}
  \frac{ds}{dt}=\sqrt{\frac{\mathcal V_t^2+\left(\frac{(f')_t\left(2-3(f')_t\right)}{f_t-\frac{1}{N}}+\frac{\left((f')_t-1\right)^2}{f_t-\mathcal P_t}\right)(\frac{1}{2}\dot{\mathcal P}_t)^2}{f_t-\frac{1}{N}}}
\end{equation}
with $\mathcal V_t=\sqrt{\mathrm{Tr}(\dot\rho_t^2)}$ and $\mathcal P_t=\mathrm{Tr}\rho_t^2$, and
the superscript $f\in\mathcal F$ denotes the dependence of $f$. In fact, under the perspective of Figure 1. (a), the QSL in Eq. (\ref{QSL}) is significant with $\forall f\in\mathcal F$ because they are all tight. That is, for any fixed initial state with the diagonal form, one can always construct the quantum channel with the form of $\mathcal E(\rho_0)=p_t\rho_0+(1-p_t)\mathbb I/N$ to drive the initial state to evolve along the geodesics, where $p_t\in[0,1]$ is monotone function satisfying $p_0=1$. The detailed proof is shown in Appendix A.

Eq. (\ref{QSL}) is one of our main results. It can be seen that the evolution speed Eq. (\ref{dsdt}) depends on the quantities $\mathcal {V}_t$ and the derivative of purity $\dot{\mathcal P}_t$. $\dot{\mathcal P}_t$ is the changing rate of the purity of the quantum state, which is relevant to the strength of decoherence \cite {PhysRevResearch.2.023299,yao2016frobenius}.   $\mathcal V_t$ is the speed or the strength of the generator of $\dot{\rho}_t$ \cite{Campaioli2019tightrobust}, which is also widely studied in QSL bounds \cite{Campaioli2019tightrobust,sun2015quantum, PhysRevLett.110.050403, PhysRevLett.111.010402}.  Suppose the dynamics of the system are described by some master equation $\dot{\rho}_t=\mathcal{L}\rho_t$, one can always convert this equation into the Liouville space as $\mathrm{Vec}(\dot{\rho}_t)=\mathrm{Vec}(\mathcal{L}\rho_t)=\mathfrak{L}\mathrm{Vec}(\rho)$ with $\mathrm{Vec}(M)$ denoting the vectorization of the matrix $M$. Thus one can easy obtain \begin{equation}\label{cost_ineq}
  \mathcal V_t=\left\vert\mathrm{Vec}(\dot\rho_t)\right\vert\leq \vert \mathfrak{L}\vert\cdot\left\vert\mathrm{Vec}(\rho_t)\right\vert\leq \left\vert \mathfrak{L}\right\vert,
\end{equation}
where $\left\vert\cdot\right\vert$ is the Fubinius norm of the matrix or the vector norm.   So $\mathcal V_t$ is bounded by the characteristics of the system dynamics.

We want to emphasize that even though the QSLT depends on the function  $f(\cdot)\in\mathcal F$, any given valid function $f(x)$ induces a valid QSLT.  Next, we will demonstrate different QSLTs based on different functions.

\textit{QSLT for unitary cases.-} The optimization in Eq. (\ref{tau_opt}) is a challenging task due to multiple potential evolution trajectories.
To gain a first insight into the QSLT, we'd like to consider the particular case of unitary evolution, in which case the bound Eq. (\ref{QSL}) can be rewritten as
\begin{equation}\label{QSL_uni}
  \tau_{uni}^f=\frac{\arccos \frac{\mathrm{Tr}\rho_0\rho_\tau-\frac{1}{N}+f-\mathrm{Tr}\rho_0^2}{f-\frac{1}{N}}}{\frac{1}{\tau}\int_0^\tau dt \sqrt{{\mathrm{Tr}\dot{\rho}_t^2}\bigl/(f-\frac{1}{N})}},
\end{equation}
where we take $f=const\geq \mathrm{Tr}\rho_0^2$, for the unitary trajectory, it only includes the states sharing the same purity, hence for this case, $f=const\geq\mathrm{Tr}\rho_0^2$ satisfy our constraints.  Maximizing  $ \tau_{uni}^f$ over $f\in\mathcal F$,  one can immediately find that the maximum is achieved for $f=\mathrm{Tr}\rho_0^2$ as \begin{equation}\label{QSL_ss}
  \tau_{uni}^{p_0}={\arccos \frac{\mathrm{Tr}\rho_0\rho_\tau-\frac{1}{N}}{\mathrm{Tr}\rho_0^2-\frac{1}{N}}}\biggl /\frac{1}{\tau}\int_0^\tau \sqrt{\frac{\mathrm{Tr}\dot{\rho}_t^2}{\mathrm{Tr}\rho_0^2-\frac{1}{N}}}.
\end{equation}
This is due to $\partial \tau_{uni}^{f}/\partial f<0$ for $f=const\geq\mathrm{Tr}\rho_0^2$, hence $\tau_{uni}^f<\tau_{uni}^{p_0}$ for the given $H$ and initial state $\rho_0$.
Eq. (\ref{QSL_ss}) is exactly the saturable bound presented in Refs. \cite{PhysRevLett.120.060409,PhysRevResearch.2.023299,H_rnedal_2023}.  Actually the above fact also implies that $f=const>\mathrm{Tr}\rho_0^2$ is not attainable for the unitary dynamics, because  a particular $\tilde{f}> \mathrm{Tr}\rho_0^2$ allowing a saturable bound will contradict with the monotonically decreasing $ \tau_{uni}^{f}$.

\textit{QSLT for general cases.-}Next, we will introduce several particular QSLTs with certain choices of the function $f(x)$.

\textit{Case 1}.-Let $f=const\equiv\alpha^2\geq 1$, one will find that the QSLT in Eq. (\ref{QSL}) becomes
\begin{equation}\label{QSL_case3}
  \tau_{qsl}^{c}=\frac{\sqrt{\alpha^2-\frac{1}{N}}\arccos\frac{\mathrm{Tr}\rho_0\rho_\tau-\frac{1}{N}+\sqrt{\alpha^2-\mathrm{Tr}\rho_0^2}\sqrt{\alpha^2-\mathrm{Tr}\rho_\tau^2}}{\alpha^2-\frac{1}{N}}}{\frac{1}{\tau}\int_0^\tau dt\sqrt{\mathrm{Tr}\dot{\rho}_t^2+\frac{1}{\alpha^2-\mathrm{Tr}\rho_t^2}(\mathrm{Tr}\rho_t\dot{\rho}_t)^2}}.
\end{equation}
 We further let $\alpha\rightarrow \infty$, we have
 \begin{equation}\label{QSL_case31}
   \tau_{qsl}^{f_1}=\frac{\sqrt{\mathrm{Tr}(\rho_0-\rho_\tau)^2}}{\frac{1}{\tau}\int_0^\tau dt\sqrt{\mathrm{Tr}\dot\rho_t^2}},
 \end{equation}
This is the QSLT bound in Ref. \cite{Campaioli2019tightrobust}.

\textit{Cases 2a}.- Let $f(x)=\frac{(1-\tilde\alpha)^2}{N}+x$ with $\tilde\alpha\in\mathbb R$ and $\tilde\alpha\neq 1/N$, then Eq. (\ref{QSL}) becomes
\begin{equation}\label{QSL_case2}
  \tau_{qsl}^{\alpha}=\frac{\arccos\frac{\mathrm{Tr}\rho_0\rho_\tau-\alpha}{\sqrt{\mathrm{Tr}\rho_0^2-\alpha}\sqrt{\mathrm{Tr}\rho_\tau^2-\alpha}}}{\frac{1}{\tau}\int_0^\tau dt\frac{\sqrt{\mathrm{Tr}(\dot\rho_t)^2(\mathrm{Tr}\rho_t^2-\alpha)-(\mathrm{Tr}\rho_t \dot\rho_t)^2}}{\mathrm{Tr}\rho_t^2-\alpha}}
\end{equation}
with $\alpha=2\tilde\alpha-\tilde\alpha^2N\in(-\infty, 1/N)$.
It should be noticed that the QSL (\ref{QSL_case31}) can also be induced by Eq. (\ref{QSL_case2}) by letting $\alpha\rightarrow \infty$ as well, and Eq. (\ref{QSL_ss}) can also be obtained for unitary evolution with $\alpha=1/N$. Besides, the QSL  (\ref{QSL_case2}) can be saturated by the dynamics with the form of $\dot\rho_t=\dot\beta_t C$, where $C$ is time-independent, and $\beta_t$ is a monotone function of $t$. The proof is shown in Appendix D.  Additionally, one will find that Eq.  (\ref{QSL_case2}) with $\alpha=0$ can also induce the QSL in our previous work \cite{PhysRevA.108.052207},  which will be explicitly given in the following Eq. (\ref{QSL_case2_1}).

\textit{Cases 2b}.- Let $f=f_2(x)=x+1/N$, i.e., $\alpha=0$ for Eq. (\ref{QSL_case2}), then Eq. (\ref{QSL_case2}) becomes
\begin{equation}\label{QSL_case2_1}
  \tau_{qsl}^{f_2}=\frac{\arccos\frac{\mathrm{Tr}\rho_0\rho_\tau}{\sqrt{\mathrm{Tr}\rho_0^2}\sqrt{\mathrm{Tr}\rho_\tau^2}}}{\frac{1}{\tau}\int_0^\tau dt \frac{\sqrt{\mathrm{Tr}\dot{\rho}_t^2\mathrm{Tr}\rho_t^2-(\mathrm{Tr}\rho_t\dot{\rho}_t)^2}}{\mathrm{Tr}\rho_t^2}},
\end{equation}
which is our attainable QSLT  in  \cite{PhysRevA.108.052207} again.

\textit{Case 3}.- Let $f=f_3=\max_{t\in[0,\tau]}\mathrm{Tr}\rho_t^2\equiv \mathrm{Tr}\tilde{\rho}^2$,  then the bound in Eq. (\ref{QSL}) can be given as
\begin{equation}\label{QSL_case1}
  \tau_{qsl}^{f_3}=\frac{\arccos\frac{\mathrm{Tr}\rho_0\rho_\tau-\frac{1}{N}+\sqrt{\mathrm{Tr}\tilde{\rho}^2-\mathrm{Tr}\rho_0^2}\sqrt{\mathrm{Tr}\tilde{\rho}^2-\mathrm{Tr}\rho_\tau^2}}{\mathrm{Tr}\tilde{\rho}^2-\frac{1}{N}}}{\frac{1}{\tau}\int_0^\tau dt\sqrt{\frac{\mathrm{Tr}(\dot\rho_t^2)+\frac{1}{\mathrm{Tr}\tilde{\rho}^2-\mathrm{Tr}\rho_t^2} \mathrm{Tr}\left(\rho_t\dot\rho_t\right)^2}{\mathrm{Tr}\tilde{\rho}^2-\frac{1}{N}}}}.
\end{equation}

If $\tilde{\rho}=\rho_0$, Eq. (\ref{QSL_case1}) will own the same numerator as $\tau_{uni}^{p_0}$. We denote the QSLT in this case by $ \tau_{qsl}^{f^0_3}$. Note that only when the dynamics is unitary,  $ \tau_{qsl}^{f^0_3}$ is completely equivalent to the bound $  \tau_{uni}^{p_0}$.  In the non-unitary process,  $ \tau_{qsl}^{f^0_3}$ and $\tau_{uni}^{p_0}$ have the same form of distance function, but they are different due to the existing term of $\left(\mathrm{Tr}\rho_t\dot\rho_t\right)^2$, which relates to the rate of change of purity respective to time $t$.

In the unitary evolution scenario, one can get $ \tau_{qsl}^{f_3}= \tau_{uni}^{p_0}$ in terms of the derivation of Eq. (\ref{QSL_ss}).  Ref.  \cite{PhysRevLett.120.060409, PhysRevResearch.2.023299} have shown that  $  \tau_{uni}^{p_0}$  is attainable for the unitary evolution subject to the  initial state $
  \rho_0=\lambda\left\vert 0\right\rangle\left\langle 0\right\vert +(1-\lambda)\left\vert 1\right\rangle\left\langle 1\right\vert$ and the Hamiltonian
  $H=e^{i\phi}\left\vert 0\right\rangle\left\langle 1\right\vert+e^{-i\phi}\left\vert 1\right\rangle\left\langle 0\right\vert$. Let $\left\vert k\right\rangle, k=0,1$ represent the eigenvectors of the initial state $\rho_0$, then for any given $\rho_0$,
the dynamics governed by the above Hamiltonian $H$ drives the initial state $\rho_0$  to evolve along the geodesics. However, the attainability is only verified in the subspace spanned by two eigenstates of $H$.

In the nonunitary scenario,   for any given initial state $\rho_0$, one can always find that the QSLT Eq. (\ref{QSL_case1}) can be saturated by the corresponding dynamics  satisfying
\begin{equation}\label{geodesics}
  \dot{\rho}_t=\dot{p}_t\left(\rho_0-\frac{\mathbb{I}}{N}\right),\\
  \rho_t=p_t\rho_0+(1-p_t)\frac{\mathbb I}{N},
\end{equation}
where $p_t\in[0,1]$ can be any monotone function satisfying $p_0=1$. Thus, the geodesics is a convex combination of any given $\rho_0$ and the maximally mixed state, namely,  the evolution trajectory passes through the maximally mixed state.  In Appendix A, we give the details of the proof.  In the following, we will give an explicit example to show the existence of these dynamics for any initial state. In this sense, the bound $  \tau_{qsl}^{f_3}$ is attainable in any dimensional for non-unitary evolution.

Based on the bound (\ref{QSL_case1}), at least in the two-dimensional case, for any given initial state, bound (\ref{QSL_case1}) can be saturated by both unitary and non-unitary dynamics. Thus, our QSLT $\tau_{qsl}^{f_3}$ provides a unified attainable bound for both unitary and non-unitary dynamics.  We emphasize the two-dimensional case in that for any given initial state in higher-dimensional cases (N>2), the bound (\ref{QSL_case1}) can only be saturated by non-unitary dynamics. Still, unitary dynamics are disabled to saturate it. Therefore, when an initial state of any dimension is given, whether an attainable QSL bound exists for the unitary dynamics in higher dimensions remains an exciting and unresolved problem. In the next section, we'll provide explicit examples for any given initial state to construct the optimal dynamics to drive the initial state to evolve along the geodesic and provide numerical examples to test the tightness.

\section{Attainablity and tightness}
\textit{Dynamical example for attainability.}-We first emphasize that the attainability of $\tau_{qsl}^{f_3}$ means for any given initial state, one can always find a corresponding dynamics to drive the state to evolve along its geodesic, which corresponds to the case shown in Figure \ref{qsl_fig} (a).  $\tau_{qsl}^{f_1}$ and $\tau_{qsl}^{f_2}$ have been analytically and numerically studied in Ref. \cite{Campaioli2019tightrobust,mai2023tight}.  It is obvious that the saturation dynamics for $\tau_{qsl}^{f_3}$ is given by Eq. (\ref{geodesics}),  which happens to be the purely depolarizing dynamics.  It describes that at the moment $t$, the state $\rho_t$ remains unchanged with a probability of $p_t$, and with a probability of $1-p_t$, it is depolarized to the maximally mixed state.

\textit{Tightness.-}As mentioned above, our proposed three QSLTs are all attainable for any given initial state in the sense of evolution along the geodesic, as shown in Figure. \ref{qsl_fig} (a).  Thus, the tightness is meaningless.  For our main results, we have given a complete story.  However, in the literature, one could always consider the concept of tightness, which actually points to a different scenario.

We have said that QSL has different understandings. Once a particular trajectory or a fixed quantum system is considered,  one always expects to find out which QSLT can be closer to the minimal evolution time, which is a different understanding of QSLT from the frame of Figure. \ref{qsl_fig} (a).  In this sense, our QSLTs can usually serve as the lower bound of the QSLT for the case of fixed trajectories or fixed initial and final states, etc., unless the considered dynamics are the saturation dynamics mentioned above.
However,   it is challenging to evaluate the tightness of a QSLT accurately. A typical example is that given initial and final states in open systems, as shown in Figure. \ref{qsl_fig} (c), the QSLT depends on the infinitely many evolution trajectories for a given pair of initial and final states. It is impossible to exhaust all the potential dynamics if there is no other limit.

From a different perspective, if one could find one QSLT is always more significant than the other for any allowed dynamics, the larger QSLT is tighter than the smaller one, even though neither is attainable.  Therefore,  evaluating a QSLT for given dynamics can at least reveal how the evolution is close (tight) to the optimal (attainable) trajectory.  So, combining different QSLT bounds can provide a tighter one if they present inconsistent forms for the same dynamics. In this sense,  our QSLT family offers a very significant result. Since our QSLT family depends on an adjustable function, one can  obtain a tight and optimal QSLT $ \tau_{qsl}^{opt}$  by maximizing all potential functions $f(\cdot)$,  i.e.,
\begin{equation}\label{tau_opt}
  \tau_{qsl}^{opt}=\max_{f\in\mathcal F}\tau_{qsl}^f.
\end{equation}
Note that if one restricts $f$ is constant, then Eq. (\ref{tau_opt}) is minimised for $f=\mathrm{Tr}\rho_0^2$.

As mentioned, the tightness of QSLTs depends on the evolution trajectories.  They're comparable only when the trajectory is given, except that they share the same form of the speed, i.e., the denominator of QSLTs. This point of view is also supported by Kavan Modi et al. \cite{Campaioli2019tightrobust}. Hence, it's impossible to conclude which QSLT derived from the family  Eq.  (\ref {QSL}) is generally tighter. Still, for the unitary dynamics, $\tau_{qsl}^{f_i}$, $i=1,2,3$ are comparable. It's reported that $\tau_{qsl}^{f_3}>\tau_{qsl}^{f_1}$ for the unitary trajectory  \cite{Campaioli2019tightrobust}, and $\tau_{qsl}^{f_2}>\tau_{qsl}^{f_1}$ holds for the pure state under purity unchanging dynamical process \cite{mai2023tight}, the continuity of QSLTs guarantees these inequalities hold for some trajectories closing to the unitary trajectories.

\begin{figure}
\includegraphics[width=0.5\textwidth]{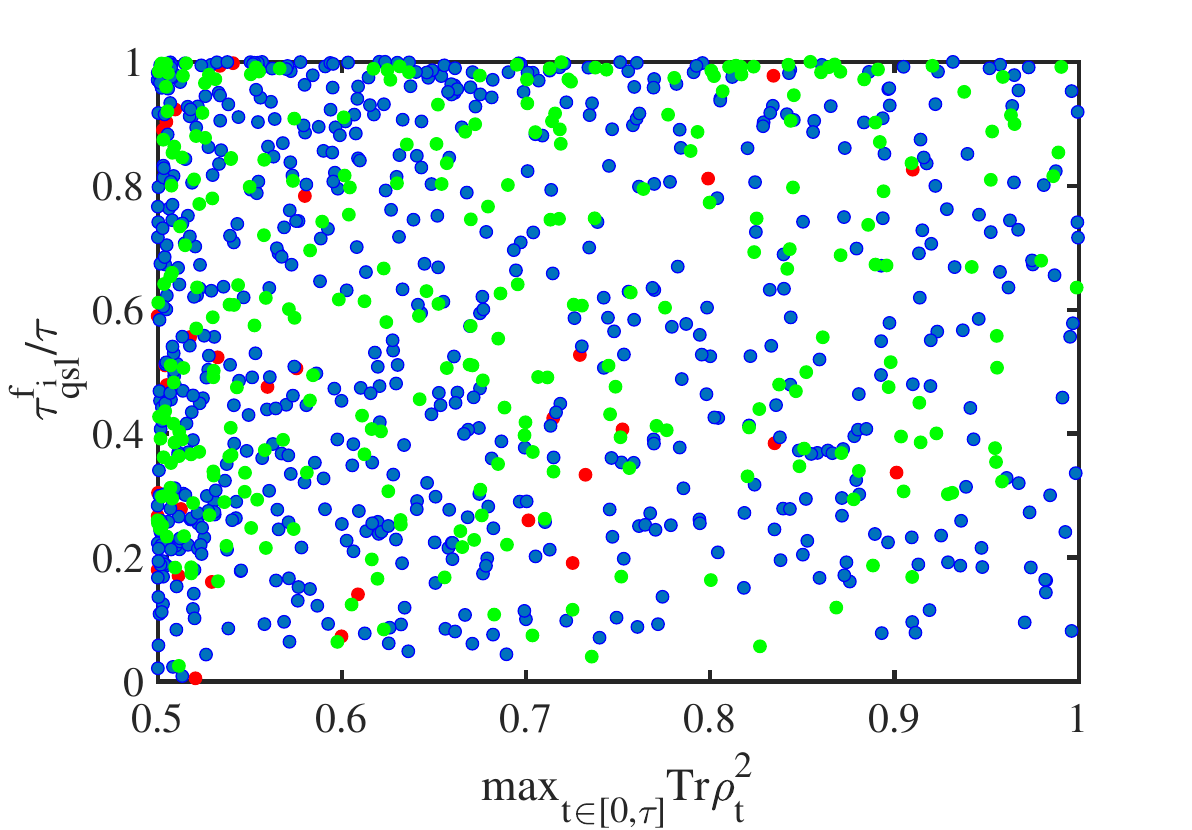}
\includegraphics[width=0.5\textwidth]{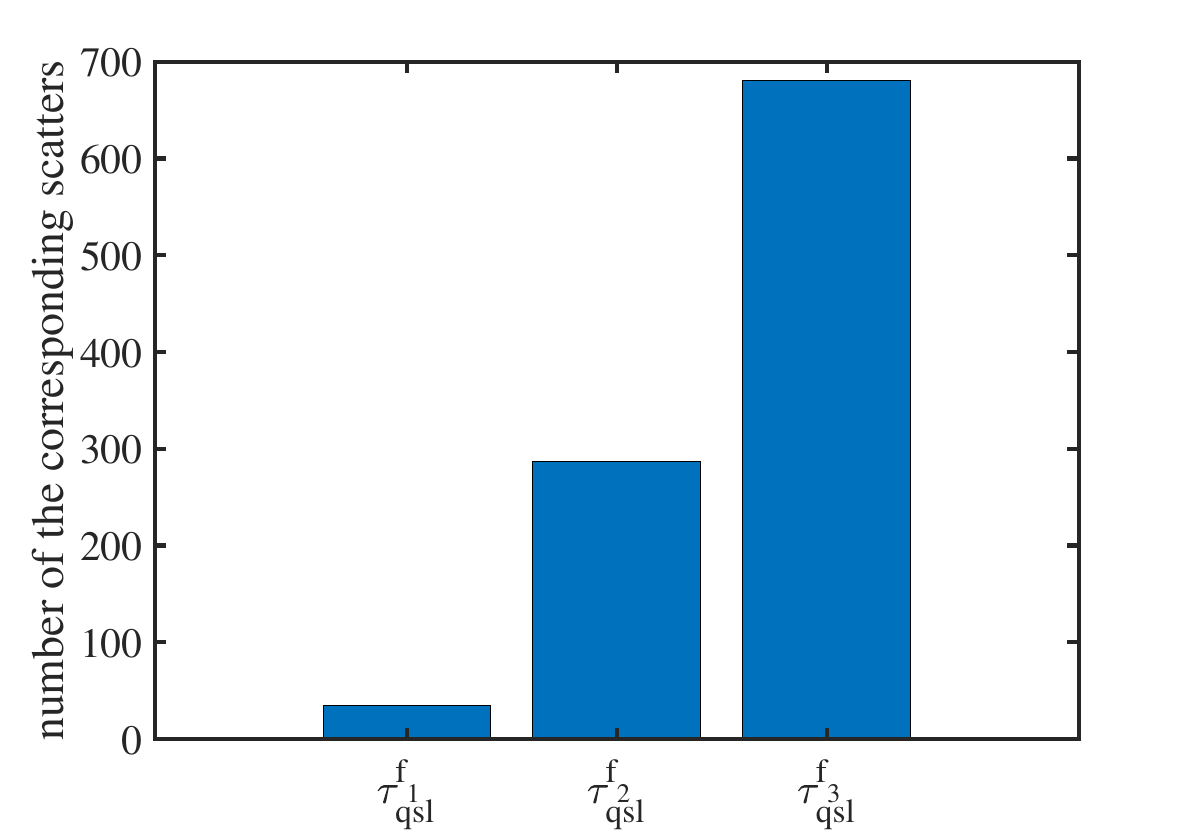}
\caption{Scatter plots of $\tau_{qsl}^{f_i}$, $i=1,2,3$ for randomly generated dynamics and initial state.  We randomly generate $1000$ dissipative dynamics to drive an initial state randomly generated as well, to evolve for a fixed duration $\tau=1$.  We plot the largest value of $\tau_{qsl}^{f_{1,2,3}}$ for each randomly dynamical process in the left figure with the red, green and blue scatters representing $\tau_{qsl}^{f_1}$, $\tau_{qsl}^{f_2}$ and $\tau_{qsl}^{f_3}$.  In the right figure, we list the numbers of each color,  which shows that $\tau_{qsl}^{f_3}$ predominates.
}\label{random_fig}
\end{figure}

To evaluate the tightness of the proposed three QSLTs, we would like to mention $\tau_{qsl}^{f_3}>\tau_{qsl}^{f_1}$ for unitary evolution \cite{Campaioli2019tightrobust},  and  $\tau_{qsl}^{f_2}>\tau_{qsl}^{f_1}$  for unitary evolution of pure states \cite{mai2023tight}. To consider general cases, we focus on a unitary evolution of a $2\otimes 2$-dimensional composite system, including the system $S$ and its environment $E$ with the initial state given by  $\rho_S\otimes \rho_E$. Thus, we can obtain a nonunitary dynamics of the system $S$  by partially tracing out the environment $E$.
We randomly generate $1000$ diagonal $4\times 4$ Hamiltonians and initial states, by which we calculate the corresponding time evolution operators and let the evolution time be a unit, i.e., $\tau=1$.
  These randomly generated qubit systems are sampled for comparing the bounds given by Eqs. (\ref{QSL_case31},\ref{QSL_case2},\ref{QSL_case1}). The result is shown in Figure. \ref{random_fig},  where each scatter point in Figure. \ref{random_fig} corresponds to the largest one of $\left\{\tau_{qsl}^{f_{1}},\tau_{qsl}^{f_{2}},\tau_{qsl}^{f_{3}}\right\}$ in single sample.  We plot the scatters red for the cases when $\tau_{qsl}^{f_1}$ is the largest one; the blue and yellow scatters are plotted for the cases similarly for $\tau_{qsl}^{f_3}$ and $\tau_{qsl}^{f_2}$, respectively. It can be seen that in Figure \ref{random_fig},  $\tau_{qsl}^{f_3}$ shows the strongest tightness for many of the considered samples, and the ratio of the number of scatters of red: yellow: blue is about $0.034:0.298:0.668$. These random samples show that the tightness of QSLTs depends on the evolution trajectory, and the tightness of the QSLT bound can be enhanced by optimizing over the function $f\in\mathcal F$ in Eq. (\ref{f}). To get QSLT with stronger tightness, we suggest the combination of QSLT as
$
    \tau_{qsl}^{comb}=\max_{i=1,2,3}\tau_{qsl}^{f_i},
$
which is just a very weak case of Eq.  (\ref{tau_opt}).

\section{Discussion and conclusion}
This paper established a family of QSLs with an adjustable function.  Different choices of the function can induce different QSLTs. We considered three functions and obtained three QSLTs.  We reveal that two of these QSLTs align precisely with those previously presented in the works,  and the third one emerges as a unified measure applicable to both open and closed quantum systems. All three QSLTs are practically achievable for any given initial quantum state in the sense that a dynamical process can drive the initial state to evolve along the corresponding geodesic path in the quantum state space.
According to the attainability of $\tau_{qsl}^{f_3}$, there are always the optimal unitary and non-unitary dynamics saturating the QSL bound for the two-dimensional case.  However, we have only found the optimal non-unitary dynamics for higher dimensional cases for any initial state.  Whether a so-called optimal unitary dynamics exists to saturate the QSLT bound remains an open problem.

The tightness relevant topic is an interesting and challenging problem on QSL.  In this sense, the QSL family provides an attractive aspect for finding tighter QSL by optimizing the adjustable function.  For example,  other optimization functions $f$ could induce the tightest bound on time for different dynamics. However, finding the optimal function $f $ is a complex functional optimization problem, and it is generally difficult to obtain an analytical solution.  Interestingly, for the unitary evolution case,  the optimization on $f$ is analytically solved, which shows the optimal one is precisely the QSLT presented in Ref. \cite{PhysRevLett.120.060409}.
Of course,  developing a tight bound in other scenarios like Figure \ref{qsl_fig} (c) is quite difficult but very significant, especially in quantum control \cite{PhysRevA.84.022305}, quantum computation, and so on. Some numerical methods \cite{Deffner_2014} are raised to find the optimal evolution path, which may work for some particular cases but is not straightforward to the general cases of unpredictable physical processes.  To sum up,  we believe that our work deepens our understanding of the QSL and opens up a new route to study QSL.

\section*{Acknowledgements}
This work was supported by the National Natural Science Foundation of China under Grants No.12175029.

\bibliographystyle{quantum}
\bibliography{reference}

\onecolumn
\appendix

\section{A saturation case of QSL $\tau_{qsl}^f$}
Now we show that dynamics Eq. (\ref{geodesics}) can saturate the bound (\ref{QSL}).  If $\rho_t$ takes the form of Eq. (\ref{geodesics}),   $F(\rho_t)$ will be a linear combination of $F(\rho_0)$ and $F(\mathbb I/N)$ due to the fact that $F(\rho)$ is a linear combination of $\rho$ and $\mathbb I/N$ for any given $\rho$. Then, one can always obtain
\begin{equation}\label{appa_1}
  F(\rho_t)=x_tF(\rho_0)+y_tF(\frac{\mathbb I}{N}),
\end{equation}
where
\begin{equation}\label{appa_2}
  y_t=-cx_t\pm\sqrt{1-(1-c^2)x_t^2}
\end{equation}
with $c=\left\langle F(\rho_0),F(\mathbb I/N)\right\rangle$. Eq. (\ref{appa_2}) is obtained by using the condition  $\left\vert F(\rho_t)\right\vert=1$. Because the distance defined as Eq. (\ref{D}) is actually the Euclidean angle of the unit vectors $F(\rho_0)$ and $F(\sigma)$,  the metric can also be expressed as
\begin{eqnarray}\label{appa_3}
ds^2=\left\vert \dot F(\rho_t)\right\vert^2=&\dot x_t^2+\dot y_t^2+2c\dot x_t\dot y_t\\
=&\dot x_t^2\frac{1-c^2}{1-(1-c^2)x_t^2},
\end{eqnarray}
where we assume that $x_t$ is monotone.  An immediate result led by the monotone $p_t$ shows
\begin{eqnarray}\label{appa_4}
  x_t=\sqrt{\frac{1-\left\langle F(\rho_t),F(\frac{\mathbb I}{N})\right\rangle^2}{1-c^2}},\\
  \frac{d}{dt}x_t=\frac{dx_t}{d\mathrm{Tr}\rho_t^2}\frac{d}{dt}\mathrm{Tr}\rho_t^2=\frac{dx_t}{d\mathrm{Tr}\rho_t^2}2p_t\dot p_t(\mathrm{Tr}\rho_0^2-\frac{1}{N}).
\end{eqnarray}
It is obvious that $dx_t/d\mathrm{Tr}\rho_t^2\geq 0$. Hence, the sign of $dx_t/dt$ is identical to $\dot p_t$. Integrating the square root of Eq. (\ref{appa_3}),  one can obtain
\begin{equation}\label{appa_5}
  \int_0^\tau ds=\left\vert \arccos\sqrt{1-c^2}x_\tau-\arccos\sqrt{1-c^2}x_0\right\vert.
\end{equation}
Finally,  we have that
\begin{eqnarray}\label{appa_6}
  D(\rho_0,\rho_\tau)=\arccos\left[(1-c^2)x_0x_\tau+\sqrt{1-(1-c^2)x_0^2}\sqrt{1-(1-c^2)x_\tau^2}\right],
\end{eqnarray}
is identical to Eq. (\ref{appa_5}),  which means  $\tau_{qsl}^f=\tau$. The proof is finished.

It is worth noting that this result is independent of the choice of function $f\in\mathcal F$, which means that the QSL family presented as Eq. (\ref{QSL}) is always attainable.

\section{The mapping $F$ is injective}
This section will show that the mapping $F$ in Eq. (\ref{f}) is injective. We first emphasize that function $f\in\mathcal F$ under the following restrictions
\begin{eqnarray}
  &\frac{d}{dx}f(x) \leq 1, \\   \label{appb_1}
  &f(x) \geq x\geq 1/N, \\   \label{appb_2}
  &\sum_{i=1}^2\left\vert f(x_i)-x_i\right\vert \neq 0.   \label{appb_3}
\end{eqnarray}

Then, it's easy to calculate that these restrictions lead to
\begin{equation}\label{appb_4}
  \frac{d}{dx}\left\langle F\left(\frac{\mathbb I}{N}\right),F(\rho)\right\rangle=\frac{d}{dx}\left\langle\frac{\mathbb I}{\sqrt{N}}, F(\rho)\right\rangle=\frac{d}{dx}\sqrt{\frac{f(x)-x}{f(x)-\frac{1}{N}}}\leq 0,
\end{equation}

where
\begin{equation}\label{appb_5}
  \frac{d}{dx}\sqrt{\frac{f(x)-x}{f(x)-\frac{1}{N}}}=\frac{(\frac{d}{dx}f(x)-1)(f(x)-\frac{1}{N})-\frac{d}{dx}f(x)(f(x)-x)}{2(f(x)-\frac{1}{N})\sqrt{f(x)-x}\sqrt{f(x)-\frac{1}{N}}}.
\end{equation}

The final inequality in Eq. (\ref{appb_4}) is obtained as follows.  From $\frac{f(x)-\frac{1}{N}}{x-\frac{1}{N}}-\frac{d}{dx}f(x)\geq 1-\frac{d}{dx}f(x)\geq 0$,
 we  have
\begin{equation}\label{appb_6}
  \left(\frac{d}{dx}f(x)-1\right)\left(f(x)-\frac{1}{N}\right)\geq \frac{d}{dx}f(x)\left(f(x)-x\right),
\end{equation}
which can immediately yield that Eq. (\ref{appb_5}) is non-positive.

Additionally, it's worth noting that, to avoid $\langle\frac{\mathbb I}{N}, F(\rho)\rangle=const$ for any input $\rho$, we must artificially impose further restriction $f(x)\neq x$ formally, which is restricted by Eq. (\ref{appb_3}) .

Now, we are ready to show that $f\in\mathcal F$ leads to the injective of $F$. Assuming there exists $\rho,\ \sigma\in\mathcal D$ satisfying $F(\rho)=F(\sigma)$, without losing generality, let $\mathrm{Tr}\rho^2\geq\mathrm{Tr}\sigma^2$. One can easily notice that $\langle F(\rho),F(\frac{\mathbb I}{N})\rangle\geq \langle F(\sigma),F(\frac{\mathbb I}{N})\rangle$ due to $\mathrm{Tr}\rho^2\geq \mathrm{Tr}\sigma^2$, where $'='$ holds only for $\mathrm{Tr}\rho^2=\mathrm{Tr}\sigma^2$. It means that $F(\rho)=F(\sigma)$ always leads to $\mathrm{Tr}\rho^2=\mathrm{Tr}\sigma^2$. According to the definition of $F$ in Eq. (\ref{f}), we have $\rho=\sigma$, hence $F$ is injective.

\section{$D$ is the distance in the Euclidean space}
In this section, we will see that the distance $D$ in Eq. (\ref{D}) satisfies the triangular inequality. In fact, for any Hermitian matrix $A=A^\dagger\neq 0$ and $B=B^\dagger\neq 0$, $d(A,B)=\arccos\frac{\mathrm{Tr}AB}{\sqrt{\mathrm{Tr}A^2}\sqrt{\mathrm{Tr}B^2}}$ is the Euclidean distance. Denote
\begin{equation}\label{appc_1}
  A=\sum_{j,k} a_{jk}\left\vert j\right\rangle\left\langle k\right\vert
\end{equation}
with $a_{jk}=x^A_{jk}+iy_{jk}^A$, where $x_{jk}^A,y_{jk}^A\in\mathbb R$. It is not difficult to calculate that $\left\langle A, B\right\rangle=\sum{j,k}\left(x_{jk}^Ax_{jk}^B+y_{jk}^Ay_{jk}^B\right)=\langle A\vert B\rangle$, where
\begin{equation}\label{appc_2}
  \vert A\rangle=\begin{pmatrix}
                   \vert x^A\rangle \\
                   \vert y^A\rangle
                 \end{pmatrix}
\end{equation}
and
\begin{equation}\label{appc_3}
  \vert x^A\rangle=\begin{pmatrix}
                     x_{11}^A \\
                     x_{12}^A \\
                     ... \\
                     x_{NN}^A
                   \end{pmatrix},   \vert y^A\rangle=\begin{pmatrix}
                     y_{11}^A \\
                     y_{12}^A \\
                     ... \\
                     y_{NN}^A
                   \end{pmatrix}.
\end{equation}
Hence, $\langle A, B\rangle$ is the inner product of two vectors $\vert A\rangle$ and $\vert B\rangle$,  that is,  $d(A, B)$ is the angle of two points onto the unit sphere within the Euclidean space, which satisfies the triangle inequality as well. It is obvious that $D(\rho,\sigma)=d\left(F(\rho),F(\sigma)\right)$, and $F(\rho)\neq 0$ for $\forall \rho\in\mathcal D$,  which can be proved as follows.   Assume $F(\rho)=0$, then we have
\begin{equation}\label{appc_4}
  \rho=(1-\sqrt{f(\mathrm{Tr}\rho^2)-\mathrm{Tr}\rho^2}\sqrt{N})\frac{\mathbb I}{N}.
\end{equation}
Tracing over both sides of Eq. (\ref{appc_4}), then the unit trace of density matrix requires
\begin{equation}\label{appc_5}
  0=\sqrt{f(\mathrm{Tr}\rho^2)-\mathrm{Tr}\rho^2}\sqrt{N}>0,
\end{equation}
where the final inequality is obtained by the property of $f$ that $f(x)>x$. It is obvious that Eq. (\ref{appc_5}) is impossible,  which means that the assumption we made is invalid, that is, $F(\rho)\neq 0$.
Hence $D$ always satisfy the triangular inequality.

\section{A saturation case of $\tau_{qsl}^{\alpha}$}
In this section, we will show that the dynamics in the form of
\begin{equation}\label{C}
  \dot\rho_t=\dot\beta_t C
\end{equation}
can saturate the $\alpha$-family QSL bound (\ref{QSL_case2}), where $\beta_t$ is the monotone function, and $C$ is the Hermitian matrix with zero-trace. The solution of Eq. (\ref{C}) is
\begin{equation}\label{C_solu}
  \rho_t=\beta_tC+\rho_0
\end{equation}
with $\beta_0=0$. To obtain the denominator of the bound (\ref{QSL_case2}), one may calculate the following items as
\begin{eqnarray}
     \mathrm{Tr}\dot\rho_t^2= & \dot\beta_t^2\mathrm{Tr}C^2,\label{Tr_geod}\\
     \mathrm{Tr}\rho_t^2=  & \beta_t^2\mathrm{Tr}C^2+\mathrm{Tr}\rho_0^2+2\beta_t\mathrm{Tr}\rho_0C,\label{Tr_geod2}\\
     \mathrm{Tr}\rho_t\dot\rho_t=  & \dot\beta_t\beta_t\mathrm{Tr}C^2+\dot\beta_t\mathrm{Tr}\rho_0C.\label{Tr_geod3}
\end{eqnarray}
Substituting Eqs. (\ref{Tr_geod},\ref{Tr_geod2},\ref{Tr_geod3}) into the denominator of the bound (\ref{QSL_case2}), one can obtain
\begin{eqnarray}\label{vt_geod}
 &\frac{\sqrt{\mathrm{Tr}(\dot\rho_t)^2(\mathrm{Tr}\rho_t^2-\alpha)-(\mathrm{Tr}\rho_t \dot\rho_t)^2}}{\mathrm{Tr}\rho_t^2-\alpha}=\left\vert\dot\beta_t\right\vert\frac{\sqrt{\mathrm{Tr}C^2\mathrm{Tr}\rho_0^2-\alpha \mathrm{Tr}C^2-\left(\mathrm{Tr}\rho_0C\right)^2}}{\beta_t^2\mathrm{Tr}C^2+\mathrm{Tr}\rho_0^2+2\beta_t\mathrm{Tr}\rho_0C-\alpha}\\
&=\left\vert\dot\beta_t\right\vert\frac{\sqrt{\mathrm{Tr}C^2\mathrm{Tr}\rho_0^2-\alpha \mathrm{Tr}C^2-\left(\mathrm{Tr}\rho_0C\right)^2}}{\beta_t^2\mathrm{Tr}C^2+\mathrm{Tr}\rho_0^2+2\beta_t\mathrm{Tr}\rho_0C-\alpha-\frac{\left(\mathrm{Tr}\rho_0C\right)^2}{\mathrm{Tr}C^2}+\frac{\left(\mathrm{Tr}\rho_0C\right)^2}{\mathrm{Tr}C^2}}=\left\vert\dot\beta_t\right\vert\frac{\sqrt{\mathrm{Tr}C^2\mathrm{Tr}\rho_0^2-\alpha \mathrm{Tr}C^2-\left(\mathrm{Tr}\rho_0C\right)^2}}{\mathrm{Tr}\rho_0^2-\alpha-\frac{\left(\mathrm{Tr}\rho_0C\right)^2}{\mathrm{Tr}C^2}+\left[\beta_t^2\mathrm{Tr}C^2+2\beta_t\mathrm{Tr}\rho_0C+\frac{\left(\mathrm{Tr}\rho_0C\right)^2}{\mathrm{Tr}C^2}\right]}\\
  &=\left\vert\dot\beta_t\right\vert\frac{\sqrt{\mathrm{Tr}C^2\mathrm{Tr}\rho_0^2-\alpha \mathrm{Tr}C^2-\left(\mathrm{Tr}\rho_0C\right)^2}}{\left(\mathrm{Tr}\rho_0^2-\alpha-\frac{\left(\mathrm{Tr}\rho_0C\right)^2}{\mathrm{Tr}C^2}\right)+\left(\beta_t\sqrt{\mathrm{Tr}C^2}+\frac{\mathrm{Tr}\rho_0C}{\sqrt{\mathrm{Tr}C^2}}\right)^2}
  =\left\vert\dot\beta_t\right\vert\frac{\sqrt{\mathrm{Tr}C^2\mathrm{Tr}\rho_0^2-\alpha \mathrm{Tr}C^2-\left(\mathrm{Tr}\rho_0C\right)^2}}{\frac{\left[\mathrm{Tr}C^2\mathrm{Tr}\rho_0^2-\alpha \mathrm{Tr}C^2-\left(\mathrm{Tr}\rho_0C\right)^2\right]+\left(\beta_t\mathrm{Tr}C^2+\mathrm{Tr}\rho_0C\right)^2}{\mathrm{Tr}C^2}}\\
  &=\left\vert\dot\beta_t\right\vert\frac{\frac{\mathrm{Tr}C^2}{\sqrt{\mathrm{Tr}C^2\mathrm{Tr}\rho_0^2-\alpha \mathrm{Tr}C^2-\left(\mathrm{Tr}\rho_0C\right)^2}}}{1+\left(\frac{\beta_t\mathrm{Tr}C^2+\mathrm{Tr}\rho_0C}{\sqrt{\mathrm{Tr}C^2\mathrm{Tr}\rho_0^2-\alpha \mathrm{Tr}C^2-\left(\mathrm{Tr}\rho_0C\right)^2}}\right)^2}
  =\frac{\left\vert\frac{d}{dt}\frac{\beta_t\mathrm{Tr}C^2+\mathrm{Tr}\rho_0C}{\sqrt{\mathrm{Tr}C^2\mathrm{Tr}\rho_0^2-\alpha \mathrm{Tr}C^2-\left(\mathrm{Tr}\rho_0C\right)^2}}\right\vert}{1+\left(\frac{\beta_t\mathrm{Tr}C^2+\mathrm{Tr}\rho_0C}{\sqrt{\mathrm{Tr}C^2\mathrm{Tr}\rho_0^2-\alpha \mathrm{Tr}C^2-\left(\mathrm{Tr}\rho_0C\right)^2}}\right)^2}.
\end{eqnarray}
Since $\beta_t$ is monotone, by integrating Eq. (\ref{vt_geod}), one can obtain
\begin{equation}\label{app}
  \frac{\tau_{qsl}^\alpha}{\tau}=\frac{\arccos\frac{\mathrm{Tr}\rho_0\left(\beta_\tau C+\rho_0\right)-\alpha}{\sqrt{\mathrm{Tr}\rho_0-\alpha}\sqrt{\mathrm{Tr}\left(\beta_\tau C+\rho_0\right)^2-\alpha}}}{\left\vert\left[\arctan\frac{\beta_\tau \mathrm{Tr}C^2+\mathrm{Tr}\rho_0C}{\sqrt{\mathrm{Tr}C^2\mathrm{Tr}\rho_0^2-\alpha \mathrm{Tr}C^2-\left(\mathrm{Tr}\rho_0C\right)^2}}\right]_{0}^\tau\right\vert},
\end{equation}
where $\left[x_t\right]_0^\tau=x_\tau-x_0$ is defined for the $t$-dependent function $x_t$. After a simple calculation, one can easily obtain that $\tau_{qsl}^\alpha=\tau$ for Eq. (\ref{app}), that is, the dynamics (\ref{C}) saturates the bound (\ref{QSL_case2}).

\end{document}